\documentclass{article}

\usepackage{arxiv}

\usepackage[utf8]{inputenc} 
\usepackage[T1]{fontenc}    
\usepackage{hyperref}       
\usepackage{url}            
\usepackage{booktabs}       
\usepackage{amsfonts}       
\usepackage{amsmath,amssymb}
\usepackage{nicefrac}       
\usepackage{microtype}      
\usepackage{graphicx}
\usepackage{doi}

\newcommand{\bit}{\begin{itemize}}
\newcommand{\eit}{\end{itemize}}
\newcommand{\be}{\begin{equation}}
\newcommand{\ee}  {\end{equation}}
\newcommand{\bd}{\begin{displaymath}}
\newcommand{\ed}{  \end{displaymath}}
\newcommand{\bc}{\begin{center}}
\newcommand{\ec}{\end{center}}

\newcommand{\hp}{\hat{p}}
\newcommand{\hT}{\hat{T}}
\newcommand{\hrho}{\hat{\rho}}
\newcommand{\hu}{\hat{u}}
\newcommand{\hv}{\hat{v}}

\newcommand{\hW}{\hat{W}}
\newcommand{\hE}{\hat{E}}
\newcommand{\hht}{\hat{t}}
\newcommand{\hx}{\hat{x}}
\newcommand{\hhl}{\hat{l}}
\newcommand{\hD}{\hat{D}}
\newcommand{\hC}{\hat{C}}
\newcommand{\hA}{\hat{A}}
\newcommand{\hB}{\hat{B}}

\newcommand{\hZ}{\hat{Z}}

\title{Modeling of thermonuclear fusion flames: Transition to detonation}

\author{ Peter V. Gordon \\
	Department of Mathematical Sciences\\
	Kent State University\\
	Kent, Ohio 44242, USA \\
	\texttt{gordon@math.kent.edu} \\
	\And
        \hspace{1mm}Leonid Kagan\thanks{Corresponding author} \\
          School of Mathematical Sciences\\
          Tel Aviv University\\
          Tel Aviv 69978, Israel\\
	\texttt{kaganleo@tauex.tau.ac.il} \\
	\AND
        \hspace{1mm}Gregory Sivashinsky \\
          School of Mathematical Sciences\\
          Tel Aviv University\\
          Tel Aviv 69978, Israel\\
	\texttt{grishas@tauex.tau.ac.il} \\
}

\renewcommand{\shorttitle}{Thermonuclear fusion flames: DDT}

\hypersetup{
pdftitle={A template for the arxiv style},
pdfsubject={q-bio.NC, q-bio.QM},
pdfauthor={David S.~Hippocampus, Elias D.~Striatum},
pdfkeywords={Supernovae explosions, white dwarfs,
thermal runaway of fast flames,
deflagration-to-detonation transition,
nucleosynthesis, thermonuclear
fusion},
}

\begin{document}
\maketitle

\begin{abstract}
  The paper is concerned with identification of the key mechanisms controlling
deflagration-to-detonation transition in stellar medium.
The issue of thermal runaway triggered by positive feedback between the
advancing flame and the flame-driven precompression is discussed in the
framework of a one-dimensional flame-folding model.
The paper is an extension of the authors' previous study dealing with the
non-stoichiometric fusion, $fuel \to products$, kinetics
(Phys.Rev.E, {\bf 103}(2021))
over physically more relevant, $fuel+fuel \to products$, kinetics.
Despite this change the runaway effect endures.
The transition occurs prior to merging of the flame with the flame-supported
precursor shock, i.e. the pretransition flame does not reach the threshold
of Chapman-Jouguet deflagration.
\end{abstract}


\section {Introduction}
Thermonuclear explosions of white-dwarf stars is a fundamental
astrophysical issue, the first principle understanding of which is still
commonly regarded as an open problem [1].
There is a general consensus that stellar explosions are manifestations of
the deflagration-to-detonation transition of an outward propagating
self-accelerating thermonuclear flame subjected to
instability/ turbulence-induced corrugations.
A similar problem arises in unconfined terrestrial flames where a positive
feedback mechanism leading to the pressure runaway has been identified [2-12].
As has been recently shown [13], there is no substantial difference between terrestrial and stellar DDT events as far as physical mechanisms are concerned.  Notwithstanding a considerable change in the equation of state and the reaction kinetics, the runaway effect survives. The present paper is an extension of the preceding study dealing with the non-stoichiometric fusion, $fuel\to products$,
kinetics [13] over a physically more relevant, $fuel+fuel\to products$,
kinetics [14,15].
In line with Ref.[13] (see also Ref.[16]), approaching the runaway point the
pretransition flame may stay perfectly subsonic, thereby challenging the view that to ensure the transition the flame should cross the threshold of Chapman-Jouguet deflagration [7].

\section{Formulation}

Except for the altered reaction rate and some technical details
(Sec.5, Appendixes A \& B), the proposed new model basically follows that of
Ref.[13], partially reproduced here for the convenience of the reader.

In the new formulation the planar thermonuclear fusion flame is sustained by
a single-step Arrhenius-type reaction rate specified as [14,15,17],

\be\label{e1}
  W=Z\rho^2C^2\exp{\left(-\sqrt[3]{\frac{T_a}{T}}\right)},
\ee
which may imitate $^{12}$C+$^{12}$C $\to$ products, a major energy-releasing
reaction in stellar flames.\\
Here $T$ is the temperature;
$T_a$, activation temperature;
$C$, mass fraction of the reactant;
$\rho$, gas density;
$Z$, reaction rate prefactor.

For dense stellar matter the caloric and thermodynamic equations of state for enthalpy $h$ and pressure $p$ are specified as,

\be\label{e2}
h=\frac{\gamma}{\gamma-1}\left(\frac{p}{\rho}\right),
\ee

\be\label{e3}
p=A\rho^{\gamma}+B\rho^{2-\gamma}T^2,
\ee
where $\gamma$ is the adiabatic index.
The caloric Eq.(\ref{e2}) is structurally similar to that of the ideal gas [18].
The thermodynamic Eq.(\ref{e3}) is a unification of classical Sommerfeld expansions pertinent to free electron gas that dominates the interior of white-dwarf stars.  For the nonrelativistic and ultrarelativistic limits $\gamma=5/3$ and
$\gamma=4/3$, respectively (see, e.g. Refs.[17, 19]).

For further discussion it is convenient to express coefficients $A$, $B$ in
terms of the initial pressure, densities, and temperature across the deflagration wave, assuming the latter to be isobaric.  Hence,

\be\label{e4}
p_0=A\rho_0^{\gamma}+B\rho_0^{2-\gamma}T_0^2\;,
\ee

and

\be\label{e5}
p_0=A\rho_p^{\gamma}+B\rho_p^{2-\gamma}T_p^2\;,
\ee
where $\rho_0,T_0,\rho_p,T_p$ are densities and temperatures far ahead and far behind the reaction zone in the planar isobaric flame, respectively; from here on the subscripts  $0,p$ stand for the fresh mixture and products respectively.

Equations (\ref{e4}) and (\ref{e5}) readily imply,

\be\label{e6}
A=\frac{p_0(\rho_p^{2-\gamma}T_p^2-\rho_0^{2-\gamma}T_0^2)}
{\rho_0^{\gamma}\rho_p^{2-\gamma}T_p^2-\rho_0^{2-\gamma}\rho_p^{\gamma}T_0^2}\;,
\ee

\be\label{e7}
B=\frac{p_0(\rho_0^{\gamma}-\rho_p^{\gamma})}
{\rho_0^{\gamma}\rho_p^{2-\gamma}T_p^2-\rho_0^{2-\gamma}\rho_p^{\gamma}T_0^2}\;.
\ee

Unlike chemical ideal gas flames, Eqs. (\ref{e3})-(\ref{e7}) allow for a significant increase
of temperature ($T_p \gg T_0$) under mild thermal expansion
($\rho_p \lesssim \rho_0$), typical of thermonuclear flames [1,7,15,20,21].  Despite this distinction, the positive feedback mechanism of ideal gas flames appears to hold also in thermonuclear flames.  This may be demonstrated even analytically adopting the Deshaies-Joulin approach [2] by considering the distinguished limit combing the large activation temperature with small Mach number while keeping their product finite (see Appendix A of Ref. [13] for details).

The enhancement of the flame speed in unconfined media is typically caused by instability  or turbulence-induced corrugations of the reaction zone.  The impact of corrugations may be accounted for even within the framework of a one-dimensional model by merely replacing the reaction rate term $W$ by $\Sigma^2 W$
with $\Sigma$ being the degree of flame front folding [2,5,6,8,9,10,12].
The $\Sigma^2$ -factor is suggested by the classical
Zeldovich--Frank-Kamenetskii theory (see Ref.[13] and Sec.5 below).

For the corrugated front, $x=f(y,t)$, evolving in the channel, $0<y<d$,
\be\label{e8}
\Sigma=\frac{1}{d}\int_0^d\sqrt{1+\left(f_y\right)^2}dy.
\ee

In the present formulation $\Sigma$ is treated as a prescribed time-independent parameter.  Note that the proposed $\Sigma$-model relates only to the deflagrative propagation and is not valid beyond the transition point.  Similar to the DDT in channels (Fig.13 of Ref.[5]) the level of wrinkling ($\Sigma$) is
expected to drop dramatically upon the transition.

   In thermonuclear flames the energy transport prevails substantially over momentum and mass transfer thus allowing to set the Prandtl number at zero and the Lewis number at infinity [20].  In suitably chosen units the set of governing equations for one-dimensional planar geometry thus reads as follows:\\

{\it \vspace*{-0.1cm}Continuity},
\be\label{e9}
\frac{\partial{\hrho}}{\partial{\hht}}+
\frac{\partial{\hrho \hu}}{\partial{\hx}}=0\;,
\ee
{\it Momentum},
\be\label{e10}
\frac{\partial{\hrho \hu}}{\partial{\hht}}+
\frac{\partial{\hrho \hu^2}}{\partial{\hx}}+
\frac{1}{\gamma}\frac{\partial{\hp}}{\partial{\hx}}=0\;,
\ee
{\it Energy},
\begin{eqnarray}
\frac{\partial{\hrho \hE}}{\partial{\hht}}+
\frac{\partial{\hrho \hu \hE}}{\partial{\hx}}+
\left(\frac{\gamma-1}{\gamma}\right)\frac{\partial{\hp \hu}}{\partial{\hx}}=\nonumber\\
\varepsilon\frac{\partial{}^2\hT}{\partial{\hx^2}}+(1-\sigma_p)\Sigma^2\hW\;,
\label{e11}
\end{eqnarray}
where, accounting for Eq.\eqref{e2},
\be\label{e12}
\hE=\frac{1}{\gamma}\left(\frac{\hp}{\hrho}\right)+\frac{1}{2}(\gamma-1)\hu^2\;,
\ee
{\it Mass fraction},
\be\label{e13}
\frac{\partial{\hrho \hC}}{\partial{\hht}}+
\frac{\partial{\hrho \hu \hC}}{\partial{\hx}}=-\Sigma^2\hW\;,
\ee
{\it Reaction rate} (see Eq.\eqref{e1}),
\be\label{e14}
\hW=\hZ\hrho^2\hC^2 \exp{\left[N_p(1-\hT^{-\frac{1}{3}})\right]}\;,
\ee
{\it Thermodynamic equation of state} (see Eqs.\eqref{e3} -- \eqref{e7}),
\be\label{e15}
\hp=\hA\hrho^{\gamma}+\hB\hrho^{2-\gamma}\hT^2,
\ee
where
\be\label{e16}
\hA=\frac{\sigma_p^{2-\gamma}-\theta_p^2}{\sigma_p^{2(1-\gamma)}-\theta_p^2}\;,
\ee

\be\label{e17}
\;\;\hB=\frac{\sigma_p^{2(1-\gamma)}(1-\sigma_p^{\gamma})}
         {\sigma_p^{2(1-\gamma)}-\theta_p^2}\;.
\ee
As may be readily checked,
\be\label{e18}
\hp(\hrho=1,\hT=1)=\hp(\hrho=\sigma_p^{-1},\hT=\theta_p)=1.
\ee

In the above equations the basic reference scales are  $\rho_p,\;T_p,\;p_0,\;
C_0,\;h_p=\gamma p_0/(\gamma-1) \rho_p,\;a_p=\sqrt{\gamma p_0/\rho_p}$,
and $U_p$-velocity of a planar isobaric
flame relative to the reaction products.
Hence,  $\hht=t/t_p,\;\hx=x/a_pt_p$, $\hu=u/a_p$, $\hrho=\rho/\rho_p$,
$\hp=p/p_0$,
$\hC=C/C_0$, $\hE=E/h_p$, $\hW=Wt_p/\rho_pC_0$,
$t_p=\lambda T_p/\rho_p h_p U_p^2$,
$\varepsilon=(U_p/a_p)^2$, $\sigma_p=\rho_p/\rho_0$,
$N_p=\sqrt[3]{T_a/T_p}$, $\theta_p=T_0/T_p$,
and $\lambda$ is the thermal conductivity assumed to be constant.

In Eq.\eqref{e14} $\hZ$ is the normalizing factor to
ensure that under isobaric conditions ( $\varepsilon \ll 1$) the scaled flame
speed relative to the burned gas approaches $\Sigma \sqrt{\varepsilon}$.
Evaluation of $\hZ$ is presented in the Appendix A.

Equations \eqref{e9}-\eqref{e17} are considered over a semi-infinite interval,
$0<\hx<\infty$.
The pertinent solution is required to meet the following initial and boundary
conditions:\\
{\it Initial conditions,}
\begin{eqnarray}
&&\hT(\hx,0)=\theta_p+(1-\theta_p)\exp(-\hx/\hhl),\nonumber \\
&&\hC(\hx,0)=1, ~\hp(\hx,0)=1, ~\hu(\hx,0)=0,\\
&&\hrho(\hx,0) \mbox{ is a positive solution of Eq. \eqref{e15}}.\nonumber
\label{e19}
\end{eqnarray}
{\it Boundary conditions,}
\begin{eqnarray}
&& \partial{\hT}(0,\hht)/\partial{\hx}=0,~
\hu(0,\hht)=0,~\hp(+\infty,\hht)=1, \nonumber\\
&&\hT(+\infty,\hht)=\theta_p,~ \hC(+\infty,\hht)=1,\\
&&\hrho(+\infty,\hht)=1/\sigma_p,~\hu(+\infty,\hht)=0. \nonumber
\label{e20}
\end{eqnarray}
The parameters employed are specified as follows:
\begin{eqnarray}
&& N_p=45,\quad \varepsilon=10^{-4}, \quad \theta_p=0.02, \quad \Sigma \geq 1,\nonumber  \\
 && \sigma_p=0.5, ~0.85, \quad  \gamma=4/3, ~5/3.
\label{e21}
\end{eqnarray}

The hot spot width $\hhl$ of Eq.(19) is chosen to initiate the deflagrative
mode.
At $\Sigma > \Sigma_{DDT}$ the latter becomes unfeasible triggering transition
to detonation.  On the whole $50\sqrt{\varepsilon}<\hhl<400\sqrt{\varepsilon}$.

In dimensional units the parameter set (21) may correspond, e.g. to
$U_p=100$km/s; $a_p=10,000$km/s; $T_0=2\cdot 10^9$K; $T_p=10^{11}$K;
$T_a=9.1\cdot 10^{15}$; $\rho_0=5\cdot 10^9$g/cm$^3$;
$\rho_0=2.5\cdot 10^9$g/cm$^3$, $4.25\cdot 10^9$g/cm$^3$,
which are quite realistic [1,7,15,20].

\section{Numerical simulations}

The computational method and numerical strategy employed in the present study
are similar to those of Refs. [8,13].
Resolution tests are discucced in the Appendix B.

Figures \ref{f1} and \ref{f2} show $\hD_f(\Sigma)$-dependencies and spatial profiles of
state variables close to the DDT point.

\begin{figure}[!h]
\centering\includegraphics[width=2.5in]{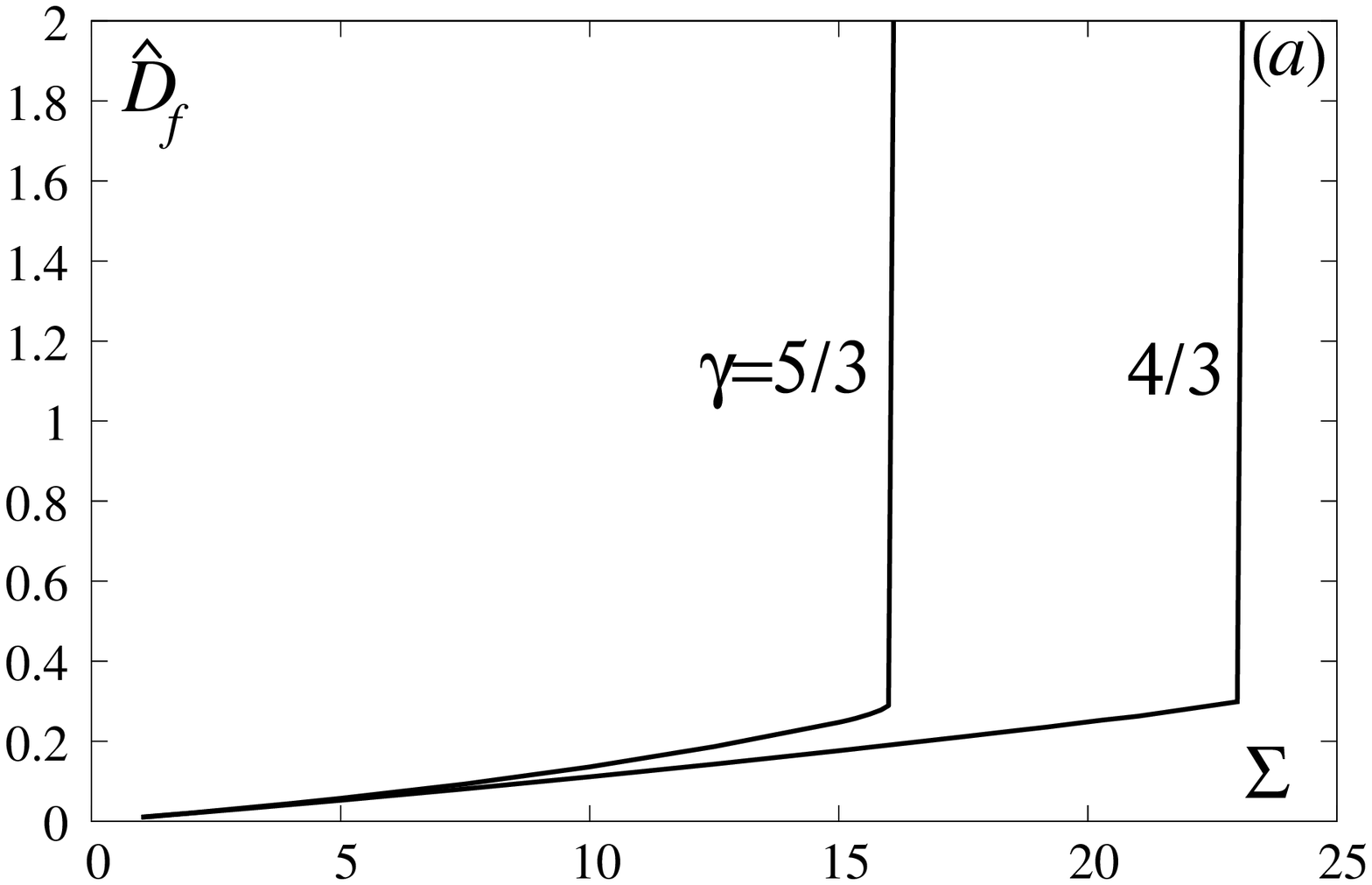}
\centering\includegraphics[width=2.5in]{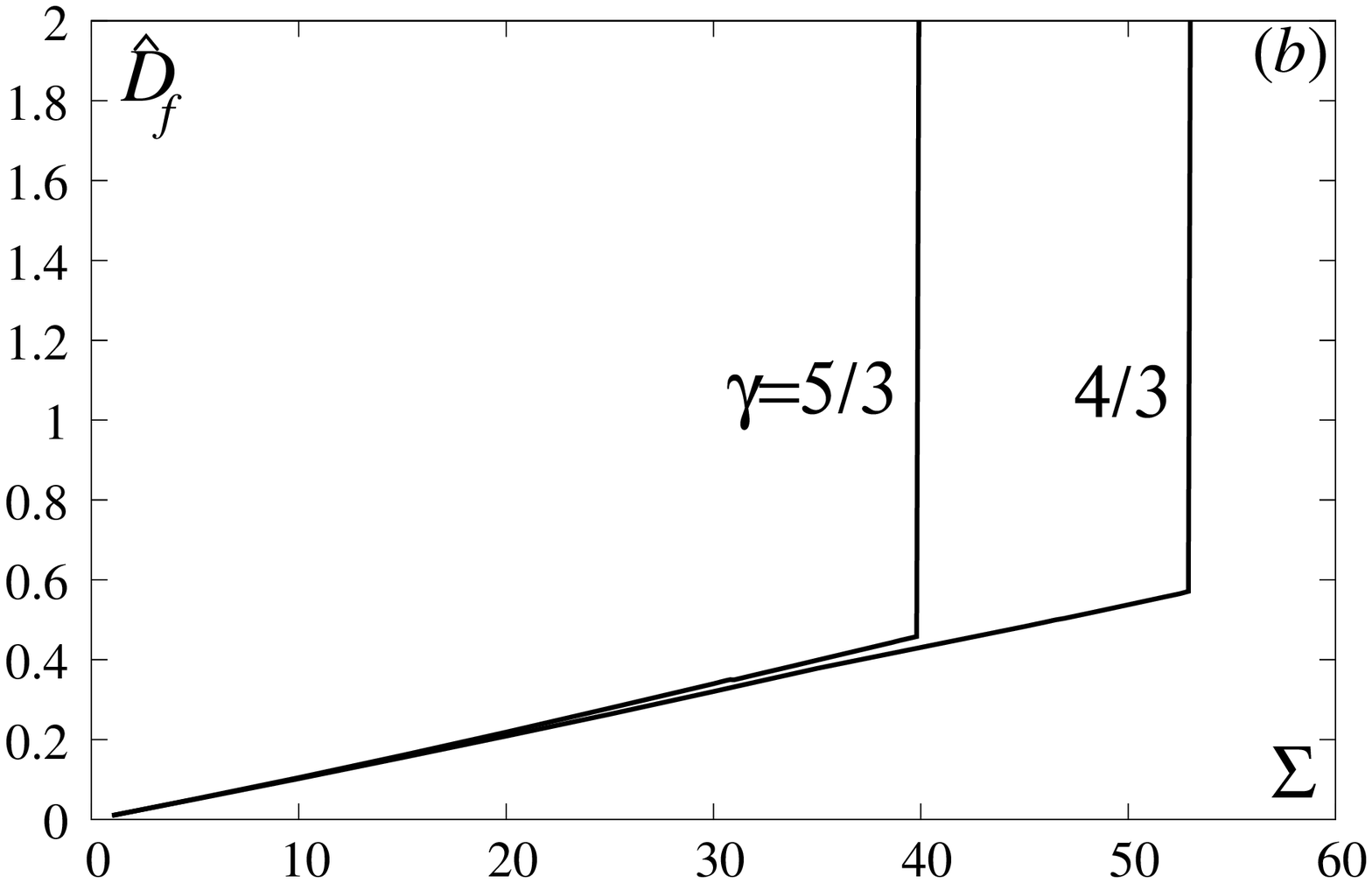}
\caption{Scaled pre-DDT flame speed $\hD_f$ vs folding factor $\Sigma$.
  Two curves on each panel correspond to $\gamma=4/3,5/3,\;\theta_p=0.02,$
$N_p=45,\;\sigma_p=0.5(a)$ and $\sigma_p=0.85(b)$}
\label{f1}
\end{figure}

According to Fig. \ref{f1}, in each case considered the flame undergoes
an abrupt
runaway when its speed reaches a critical level.
The transition invariably occurs at $\hD_f<1$, i.e. below the threshold of the
CJ-deflagration, $\hD_f=1$.

\begin{figure}[!h]
\centering\includegraphics[width=5in]{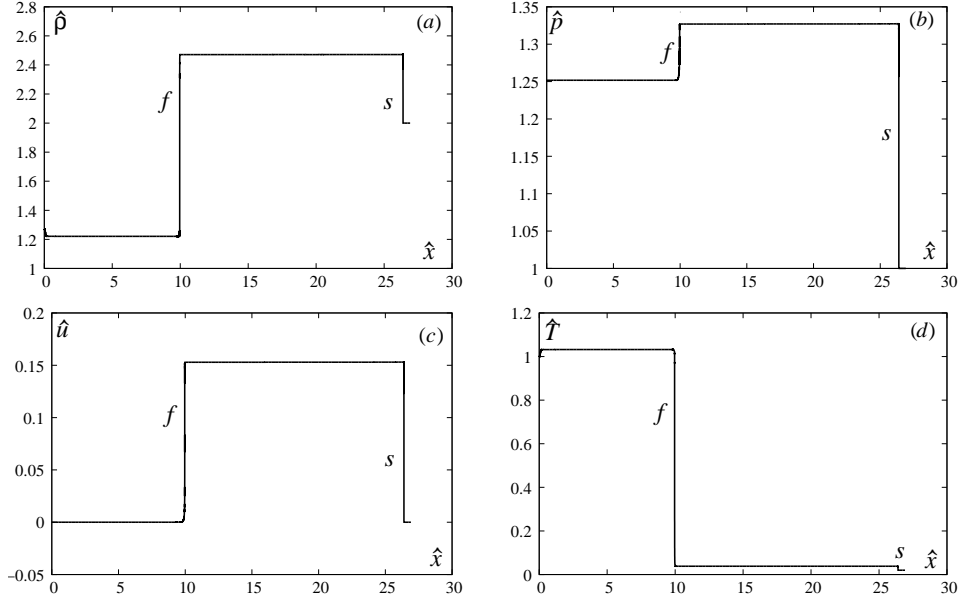}
\caption{Spatial profiles of density ($a$), pressure ($b$),
  gas velocity ($c$), and temperature ($d$) adjacent to the DDT point.
  Labels $f$ and $s$ mark the flame front and the precursor shock
  ($\gamma=4/3,\;\sigma_p=0.5,\;\theta_p=0.02,\;N_p=45,\;\Sigma=23$).
  Similar profiles for other cases of Figure \ref{f1} are not shown. 
}
\label{f2}
\end{figure}

The profiles of Fig. \ref{f2} are quite in line with what is expected for a subsonic
deflagration propagating from the channel's closed end [22].  

\section{Traveling wave solution}

Behind the precursor shock the well-settled flame assumes the form of a self-similar traveling wave (Fig. 2) whose structure may be described by a single first-order ordinary differential equation (ODE) for $\hT(\hC)$ (see Ref. [13] for details),
\begin{eqnarray}
  \left(\hrho_1 \hv_1\right)^2 \left[\frac{\hp(\hT)}{\hrho(\hT)}+
    \frac{1}{2}(\gamma-1)\hv^2(\hT)+
  (1-\sigma_p)\hC\right]+
\varepsilon \Sigma^2\hW \frac{d\hT}{d\hC}\nonumber\\
=\left(\hrho_1 \hv_1\right)^2 \left[\frac{\hp_1}{\hrho_1}+\frac{1}{2}(\gamma-1)\hv_1^2+
  (1-\sigma_p)\right],
\label{e22}
\end{eqnarray}

where $\hp(\hT)$, $\hrho(\hT)$ and $\hv(\hT)=\hD_f-\hu(\hT)$ are defined by
Eq. \eqref{e15} and relations,
\be\label{e23}
\hrho \hv = \hrho_1 \hv_1,
\ee

\be\label{e24}
\frac{\hrho^2_1 \hv^2_1}{\hrho}+\frac{1}{\gamma}\hp(\hrho,\hT)=
\hrho_1 \hv^2_1+\frac{1}{\gamma}\hp_1.
\ee

Equation \eqref{e22} is considered jointly with boundary conditions
\be\label{e25}
\hT(\hC=1)=\hT_{ign},\;\;\hT(\hC=0)=\hT_2.
\ee

In numerical simulations the ignition temperature is set as
$\hT_{ign}=1.01 \hT_1$.\\
Other relevant parameters are identical to those of Sec.2.\\  
Parameters $\hp_1,\;\hrho_1,\;\hv_1=\hD_f-\hu_1,\;\hT_1,\;\hT_2,\;\hD_f$ may be
expressed in terms of the precursor shock velocity $\hD_s$ by employing
conventional Rayleigh and Rankine-Hugoniot relations across the shock and the flame front (see Ref.[13]), augmented with the equation of state \eqref{e15}.

The problem \eqref{e22} \eqref{e25} is clearly overdetermined which allows evaluation of      
$\Sigma(\hD_f)$.

Straightforward computations show that Eq. (22) considered jointly with the
first boundary condition (25) produces a family of solutions parametrized by
$\hD_f$.
This family is monotone increasing with respect to $\Sigma$.
This observation and standard shooting arguments allow to conclude that there
exists a unique value of $\Sigma$ for which the system (22)(25) admits
a solution.

Figure 3 displays emerging $\Sigma(\hD_f)$ -dependencies.
As is readily seen the traveling wave solution ceases to exist above
$\Sigma_{max}$, which invariably falls at $\hD_f<1$, i.e., below the
CJ-deflagration point.

Similar to the situation in Ref.[13] only part of the
$\Sigma(\hD_f)$-dependency appears to be dynamically feasible.
The transition to detonation actually occurs at $\Sigma_{DDT} < \Sigma_{max}$
(Figs.1 and 4).  The traveling wave solution pertaining to
$\Sigma_{DDT} <\Sigma < \Sigma_{max}$ transpires to be unstable yielding an abrupt
transition to CJ-detonation (cf. Ref.[13]).

\begin{figure}[!h]
\centering\includegraphics[width=3.75in]{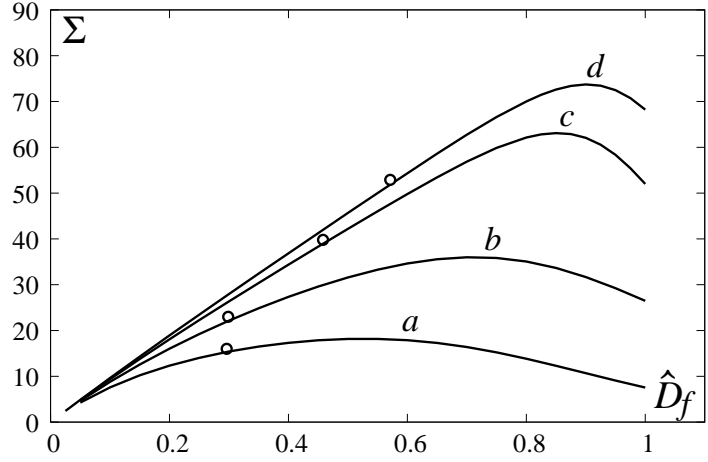}
  \caption{Folding factor $\Sigma$ vs scaled flame speed $\hD_f$, corresponding
    to $\gamma=5/3,\;\sigma_p=0.5(a);\;\gamma=4/3,\;\sigma_p=0.5(b);\;
    \gamma=5/3,\;\sigma_p=0.85(c);\;\gamma=4/3,\;\sigma_p=0.85(d)$.
    Open circles mark the DDT points of Figure 1.
}
\label{f3}
\end{figure}

\begin{figure}[!h]
\centering\includegraphics[width=3.75in]{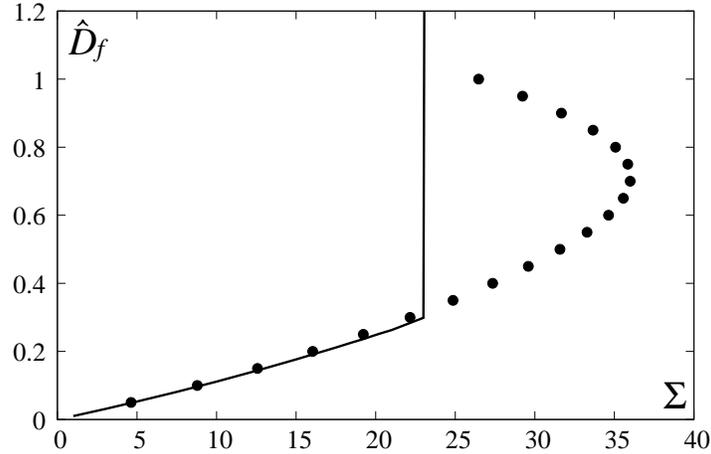}
  \caption{Illustrating the relation between the travelling wave solution (dots)
    and its dynamical counterpart (Figures 1 and 3)
    ($\gamma=4/3,\;\sigma_p=0.5,\;\theta_p=0.02,\;N_p=45$).
}
\label{f4}
\end{figure}

Figure 5 displays evolution of the reaction wave velocity $\hD_f$ emanating from
the traveling wave solution corresponding to $\Sigma=27.35$ and $\hD_f=0.4$.
The incipient dynamics, upon the oscillatory deflagrative mode,
abruptly converts into overdriven detonation, $\hD_f > \hD_{CJ}$, which
eventually evolves into the Chapman-Jouguet detonation with $\hD_{CJ}=1.805$
(see Eq. (36) of Ref. [13]).

\begin{figure}[!h]
\centering\includegraphics[width=3.75in]{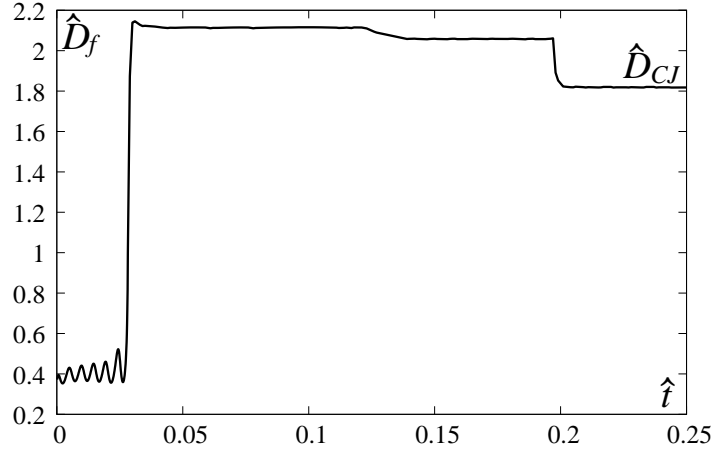}
\caption{Time record of the reaction wave velocity $\hD_f$. The initial
  conditions employed are the traveling wave profiles corresponding to
  $\gamma=4/3,\;\sigma_p=0.5,\;\theta_p=0.02,\;N_p=45,\;\Sigma=27.35$.
}
\label{f4a}
\end{figure}

\section {Concluding remarks}

The $\Sigma^2$-factor in the reaction rate term is suggested by the
Zeldovich--Frank-Kamenetskii theory valid for low Mach numbers, $\hD_f \ll 1$
[23].  For general Mach numbers the model is an extrapolation, expected to provide a reasonably good description of the physics involved.
It is certainly satisfying that transition to detonation occurs at flame
speeds $\hD_f$, considerably below unity (see Fig. 3), which could not be
foreseen in advance.
Note that in the ideal gas chemical flames, e.g. at Le$=1$, Pr$=0.75$,
$\varepsilon=0.0025,\;\sigma_p=0.125,\;N_p=5$,
depending on the reaction rate pressure dependency, 
the transition may occur either at  $\hD_f<1$ or at $\hD_f>1$ [8,9].

While the one-dimensional $\Sigma$-model is helpful for exposing the precompression-induced runaway, it conceals the fine multidimensional structure of the flame-flow interaction.  In the context of unconfined ideal gas chemical flames the latter issue has recently been addressed in Ref. [11] reproducing both the Darrieus-Landau (DL) wrinkling and, most importantly, DDT.

For the stellar medium however, due to the enormous disparity between the spatial scales involved, modeling and simulations of unconfined hydrodynamically unstable flames from first principles, while resolving all relevant scales, is not feasible either now or in the foreseeable future.  Yet, rational development and exploration of appropriately designed reduced models (accounting for the principal physics involved) is not out of reach and is expected to be quite educational [10].

\begin{figure}[!h]
\centering\includegraphics[width=3.75in]{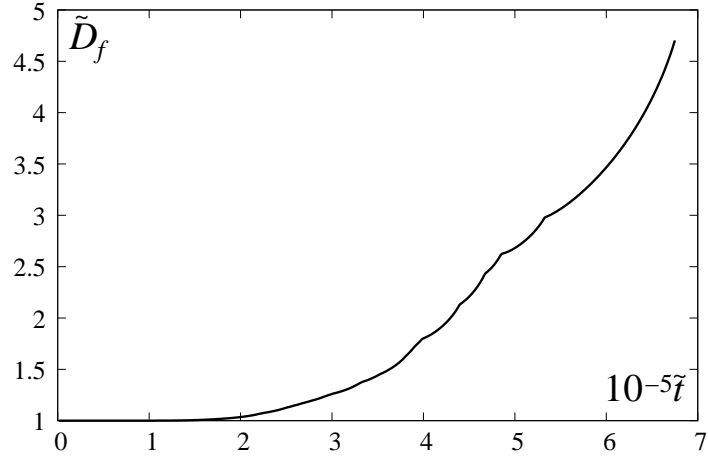}
\caption{Scaled angle-averaged flame speed $\tilde{D}_f=\overline{R}_t$ vs.
  scaled
  time $\tilde{t}$ at $G=0.0002$. The reference scales employed are $l_M$
  - the Markstein length, and $U_p$ - velocity of a planar isobaric flame
  relative to the reaction products, $G$ is the buoyancy parameter [10].
  At $l_M=1$cm, $U_p=10^6$cm/s, $\tilde{D}_{f,max}$ corresponds to
  $4.75\cdot10^6$cm/s and $\tilde{t}_{max}$ to $0.68$s.		.
}
\label{f5}
\end{figure}

\begin{figure}[!h]
\centering\includegraphics[width=3.75in]{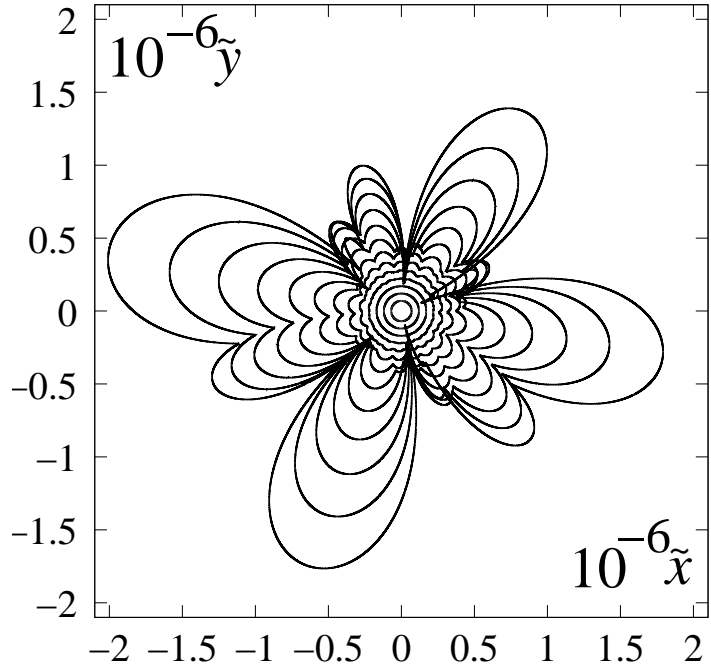}
\caption{Flame front configurations for the weakly nonlinear model [10] prior
  to the DDT event at $G=0.0002$. The reference scales employed are identical
  to those of Figure \ref{f5}; $\tilde{R}_{max}=2\cdot 10^6$ corresponds
  to $20$km.
}
\label{f6}
\end{figure}

Unlike ideal gas chemical flames, in thermonuclear flames the thermal expansion of reaction products is relatively small (Sec. 2) which justifies utilization of the Boussinesq distinguished limit [24].  The Boussinesq quasi-constant-density approximation, in turn suppresses the DL-instability whose impact is generally deemed inferior to that of the Rayleigh-Taylor (RT).  For a weak RT-instability the structure of the evolving flame becomes both quasi-planar and quasi-steady which allows reduction of the effective dimensionality of the problem.  As a result one ends up with a weakly nonlinear equation for the flame front evolution amenable to straightforward numerical simulations [10].  The weakly nonlinear model is certainly unable to capture the full morphology of the RT-mushrooming [24].
Yet the model proves adequate enough to imitate the buoyancy-induced
corrugations (Figs. \ref{f5},\ref{f6}), the inverse cascade, self-acceleration
of the front, and occurrence of the deflagrability threshold -- the precursor
of DDT.

\section* {Appendix A: Evaluation of the normalizing factor
 {\Large$\hZ$} of Eq.{\large\eqref{e14}}}
\renewcommand{\theequation}{A.\arabic{equation}}
\setcounter{equation}{0}
To evaluate $\hZ$ we turn to the traveling wave solution of Sec.4.
For the isobaric limit ($\hp=1$) the term $\hv^2$ becomes negligibly small,
$\hrho_1=1/\sigma_p$
and Eq. \eqref{e22} simplifies to,
\be\label{ea1}
  \frac{d\hT}{d\hC}=-\frac{(\hrho_1 \hv_1)^2}{\varepsilon \Sigma^2 \hW}
  \left[\frac{1-\hrho(\hT)}{\hrho(\hT)}+(1-\sigma_p)\hC\right],
\ee
where
\be\label{ea2}
\hA\hrho^{\gamma}+\hB\hrho^{2-\gamma}\hT^2=1
\ee
As mentioned in Sec.2, the normalizing factor $\hZ$ is chosen to meet the
condition,
\be\label{ea3}
\hrho_1 \hv_1=\Sigma\sqrt{\varepsilon}
\ee
Equation \eqref{ea1} then assumes a form not involving parameters
$\varepsilon$ and $\Sigma$,
\be\label{ea4}
  \frac{d\hT}{d\hC}=-\frac{1}{\hZ \hrho^2(\hT) \hC^2}
  \left[\frac{1-\hrho(\hT)}{\hrho(\hT)}+(1-\sigma_p)\hC\right]
  \exp[-N_p(1-\hT^{-\frac{1}{3}})].
\ee
Equation \eqref{ea4} should be considered jointly with boundary conditions,
\be\label{ea5}
\hT(\hC=1)=\hT_{ign},\;\;\;\hT(\hC=0)=1.
\ee

\begin{figure}[!h]
  \centering
  \includegraphics[width=2.5in]{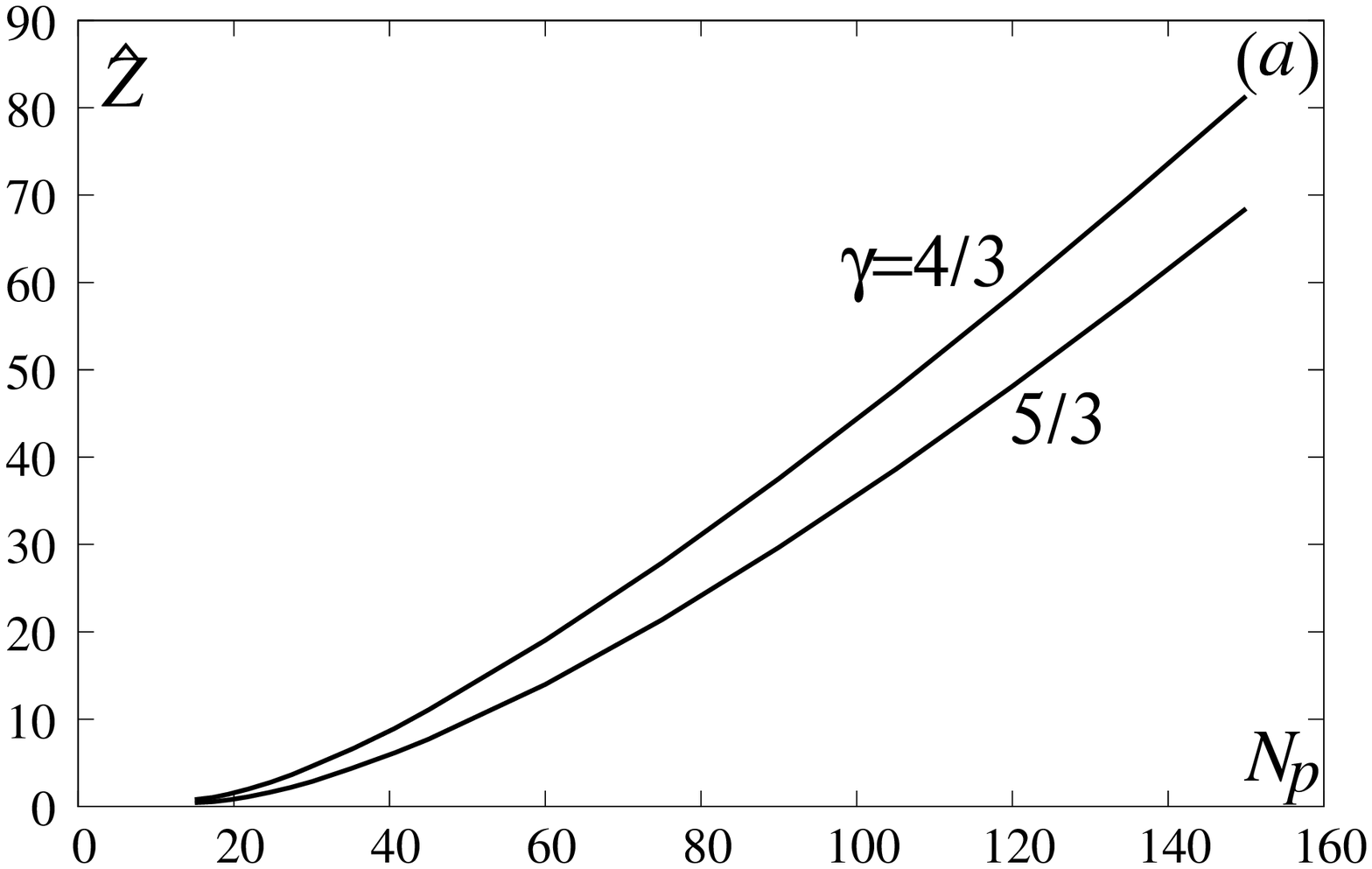}
  \includegraphics[width=2.5in]{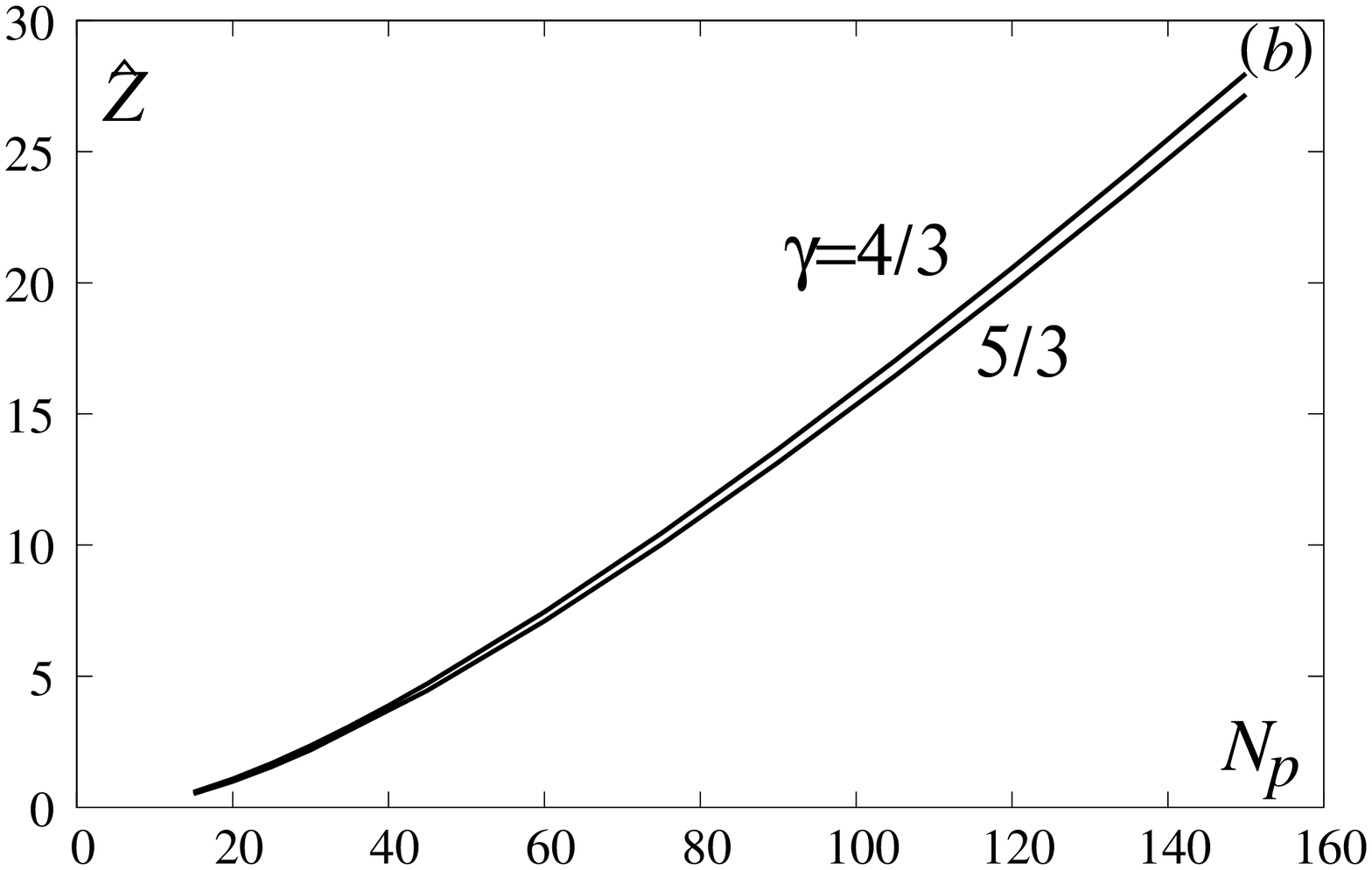}
  \caption{Normalizing factor $\hZ$ vs. $N_p$ for $15<N_p<150$,
    $\gamma=4/3,5/3,\;\theta_p=0.02,\;\sigma_p=0.5(a),\;\sigma_p=0.85(b)$
}
\label{f7}
\end{figure}

The problem \eqref{ea4} \eqref{ea5}, accounting for \eqref{ea2}, is clearly
overdetermined which allows evaluation of
$\hZ$ vs $N_p$ -dependencies calculated
for $15<N_p<150,\;\sigma_p=0.5,0.85,\;\theta=0.02,\;T_{ign}=1.01\theta_p,\;
\gamma=4/3,5/3$ (Fig. \ref{f7}).
Equation (A.4) is solved numerically with the left part of (A.5) serving as
the initial condition.
Parameter $\hZ$ is then determined by the bisection method.

Here $\hZ(N_p=45,\gamma=4/3,\sigma_p=0.5)=11.181$,
$\hZ(N_p=45,\gamma=5/3,\sigma_p=0.5)=7.693$,
$\hZ(N_p=45,\gamma=4/3,\sigma_p=0.85)=4.71829$,
$\hZ(N_p=45,\gamma=5/3,\sigma_p=0.85)=4.45626$.

\section* {Appendix B: Localizing the runaway-point and resolution tests}
\renewcommand{\theequation}{B.\arabic{equation}}
\setcounter{equation}{0}

Localizing the runaway-point is similar to that of Ref. [13].
The procedure is repeated for several spatial steps
$\Delta\hx_i=\Delta\hx_{i-1}/2$.
The results obtained are shown in Table 1.  The last three lines are used for
estimation of the convergence order (see Ref. [13]).  For $\gamma=4/3$
and $\gamma=5/3$ 
the convergence orders are 1 and 0.78, predicting $\Sigma_{DDT}^0=55.7$
and $41.5$ at $\Delta x \to 0$, respectively.\\

\begin{table}[!h]
  \caption{Folding factor $\Sigma_{DDT}$ at $\sigma_p=0.85$ under different
    resolutions.}
\label{t1}
\bc
\begin{tabular}{lll}
\hline
$\Delta x$ & $\gamma=4/3$ & $\gamma=5/3$ \\
\hline
0.0005	      & 40   & 30.8   \\
0.00025	      & 46.5 & 35.4   \\
0.000125      & 51.7 & 38.6   \\
0.0000625     & 52.5 & 39.8   \\
0.000031255   & 52.9 & 40.5   \\
\hline
\end{tabular}
\ec
\vspace*{-4pt}
\end{table}

For the smallest spatial steps $\Sigma_{DDT}$ are close enough to
$\Sigma_{DDT}^0$.
The resolutions employed are therefore quite tolerable.

\section*{Acknowledgements}

The work of L.K. and G.S. was partially supported by the Israel Science
Foundation (Grant 335/18). The work of P.V.G. was partially supported by 
the Simons Foundation (Grant 317882).
The numerical simulations were performed at the Ohio Supercomputer Center
(Grant PBS 0293-1) and the Computer Center of Tel Aviv University.



\end{document}